\documentclass[twocolumn,aps,pra]{revtex4-1}
\usepackage{epsfig}
\usepackage[english]{babel}
\usepackage{latexsym}
\usepackage{subfigure}
\usepackage{graphics}
\usepackage{epstopdf}
\usepackage{dcolumn}
\usepackage{amsmath}
\usepackage{hyperref}
\usepackage{amssymb}
\usepackage{appendix}
\usepackage{color}
\usepackage{lineno}
\usepackage{soul}
\usepackage{booktabs}
\usepackage{longtable}
\usepackage{multirow}
\usepackage{array}
\usepackage{ulem}

\begin{document}

\title{Attosecond transient absorption spectroscopy in monolayer hexagonal boron nitride}
\author{Jiayu Yan, Chenkai Zhu, Rongxiang Zhang, Xiaohui Zhao$^{\dag}$, Fulong Dong$^{*}$}

\date{\today}

\begin{abstract}
We simulate the attosecond transient absorption spectroscopy (ATAS) of monolayer hexagonal boron nitride (hBN) using the time-dependent density functional theory and two-band density-matrix equations within the tight-binding approximation.
The simulation results from the two methods are qualitatively consistent.
We focus on the fishbone structure around the gap energy of the $\textrm{M}$ point, which exhibits a temporal period equal to that of the pump laser.
To gain deeper insight into this structure, we simplify the two-band model to a single-electron model located at the $\textsc{M}$ point, allowing us to derive an analytical expression that can qualitatively reproduce the numerical results.
By isolating the influence of the Berry connection on the ATAS, our analytical results reveal that both the interband transition dipole moments and the Berry connection play important roles in the fishbone structure of the ATAS.
Moreover, we also have investigated the dependence of ATAS on the gap energy based the tight-binding approximation.
The results demonstrate that the ATAS intensity is enhanced as the gap energy increases, in agreement with our analytical prediction.
Our study may shed light on the generation mechanism of the fishbone structure of the ATAS in hBN.

\end{abstract}
\affiliation{College of Physics Science and Technology, Hebei University, Baoding 071002, China}

\maketitle

\section{Introduction}

Attosecond transient absorption spectroscopy (ATAS) is a powerful and versatile technique for probing ultrafast electronic dynamics on the attosecond timescale \cite{Eleftherios,MetteBGaarde,AnneliseRBeck,MengxiWu}.
By combining an extremely short duration of attosecond light pulses with a strong pump laser field, ATAS provides an all-optical method to investigate light-matter interactions with simultaneous high temporal and spectral resolution.
The ATAS has been successfully applied to study the electron dynamics of atoms and molecules \cite{HeWang,MHoller,ZQYang,ShaohaoChen,MichaelChini,PPeng1,PPeng2}.

In recent years, significant progress has been made in applying ATAS to condensed matter systems, including bulk solids \cite{MartinSchultze,MLucchini,FSchlaepfer,MVolkov,TOtobe,MatteoLucchini} and various two-dimensional materials \cite{KUchida,GioCistaro, Dong4}.
In periodic materials, an intriguing structure, referred to as the ``fishbone structure", has been observed in the ATAS \cite{MLucchini,TOtobe1,SYamada,Dong4}.
This characteristic spectral pattern reflects the complex interplay between the electronic band structure and the transient optical response of the system.

More recently, analytical studies on the ATAS of graphene \cite{GioCistaro,Dong4} have revealed that the fishbone structure is closely related to the band structure, particularly to the effective mass of electrons at van Hove singularity.
These studies indicate that in crystals, the ATAS are dominated by intraband electron dynamics, called by ``dynamical Franz-Keldysh effect" \cite{MLucchini,TOtobe1}.
This leads to the hypothesis that, in symmetry-broken crystals, the Berry connection may also play a significant role in intraband electron dynamics, thereby influencing the structure of the ATAS.
Moreover, previous studies have primarily considered electron excitation from core levels to conduction bands driven by the probe pulse \cite{MartinSchultze,MLucchini,GioCistaro,Dong4}.
In this context, the influence of interband transition dipole moments (TDMs) on the ATAS has also been largely overlooked.

Monolayer hexagonal boron nitride (hBN), a symmetry-broken two-dimensional crystal with a lattice structure similar to that of graphene, has recently attracted significant attention in the study of ultrafast phenomena \cite{REFSilva,Dong6,VPervak}.
It may become an ideal platform for exploring the influence of the Berry connection on the fishbone structure observed in the ATAS.
Moreover, by employing a two-band model of gapped graphene based on the tight-binding approximation, which can qualitatively describe the electric structure of monolayer hBN, the affect of interband TDMs on the ATAS can also be explored.

In this work, we investigate the ATAS of monolayer hBN by employing the time-dependent density functional theory (TDDFT) and the two-band density-matrix equations (TBDMEs).
Both methods reveal a distinctive fishbone feature near the energy gap at the $\textrm{M}$ point, with a modulation frequency equal to that of the pump laser.
This behavior contrasts with similar spectral structures reported in previous studies \cite{Dong4,MVolkovSato,SYamadaYabana,YKim}, where the modulation frequency is twice that of the pump laser.
To gain further insight into the origin of the fishbone structure, we simplify the two-band model to a single-electron model located at the $\textsc{M}$ point, from which we derive analytical expressions for the fishbone structure of the ATAS.
The analytical results can qualitatively reproduce the numerical simulations and reveal that both the interband TDMs and the Berry connection play essential roles in the fishbone structure of the ATAS.
Furthermore, by isolating the influence of the Berry connection in the ATAS, we obtain the spectra that are respectively dominated by the interband TDMs and the Berry connection, respectively.
In addition, we investigate the dependence of the ATAS on the gap energy, and reveal a qualitative consistency between our analytical predictions and numerical simulations.

This paper is organized as follows.
We describe our numerical simulation methods and results of the ATAS in Sec. \ref{s2}.
We present the analytical results based on the single-electron model in Sec. \ref{s3}.
Finally, Sec. \ref{s4} presents our conclusion.
Throughout the paper, atomic units are used if not specified.

\section{Numerical simulation of the ATAS}

\label{s2}
\subsection{Two-band density-matrix equations under tight-binding approximation}

Monolayer hBN is a two-dimensional material consisting of boron and nitrogen atoms arranged in a hexagonal lattice, as shown in Fig. 1(a).
To investigate the ATAS in monolayer hBN, we employ a tight-binding model of gapped graphene, which shares the same reciprocal lattice as hBN, as illustrated in Fig. 1(b).
Utilizing Bloch states as the basis set, the Hamiltonian $H_{0}$ takes the form
$
H_{0}=\left(\begin{array}{cc}
\Delta_g / 2 & \gamma_{0} f(\mathbf{k}) \\
\gamma_{0} f^{*}(\mathbf{k}) &- \Delta_g / 2
\end{array}\right),$
where electrons can only hop to nearest-neighbor atoms with hopping energy $\gamma_{0}=0.1$ a.u.
The function $f(\textbf{k})$ is given by $f(\textbf{k})=e^{i \texttt{k}_{x}d}+2\cos( \sqrt{3}\texttt{k}_{y}d / 2)e^{-i\texttt{k}_{x}d/2}$, where $d= 1.42$ \AA $ $ is the nearest-neighbor distance.
Diagonalizing the Hamiltonian $H_{0}$ yields the energy eigenvalues that define the dispersion relation of the conduction ($c$) and valence ($v$) bands $\varepsilon_{c}(\textbf{k})=-\varepsilon_{v}(\textbf{k})=\sqrt{\gamma^2_0|f(\textbf{k})|^2+\Delta_g^{2} / 4}$, as shown in Fig. 1(c).
Here, $\Delta_g = 0.17$ a.u. is chosen to match the energy gap at $\textrm{K}$ point with the results obtained from density functional theory (DFT) calculations.


\begin{figure}[t]
\begin{center}
{\includegraphics[width=8.5cm,height=8cm]{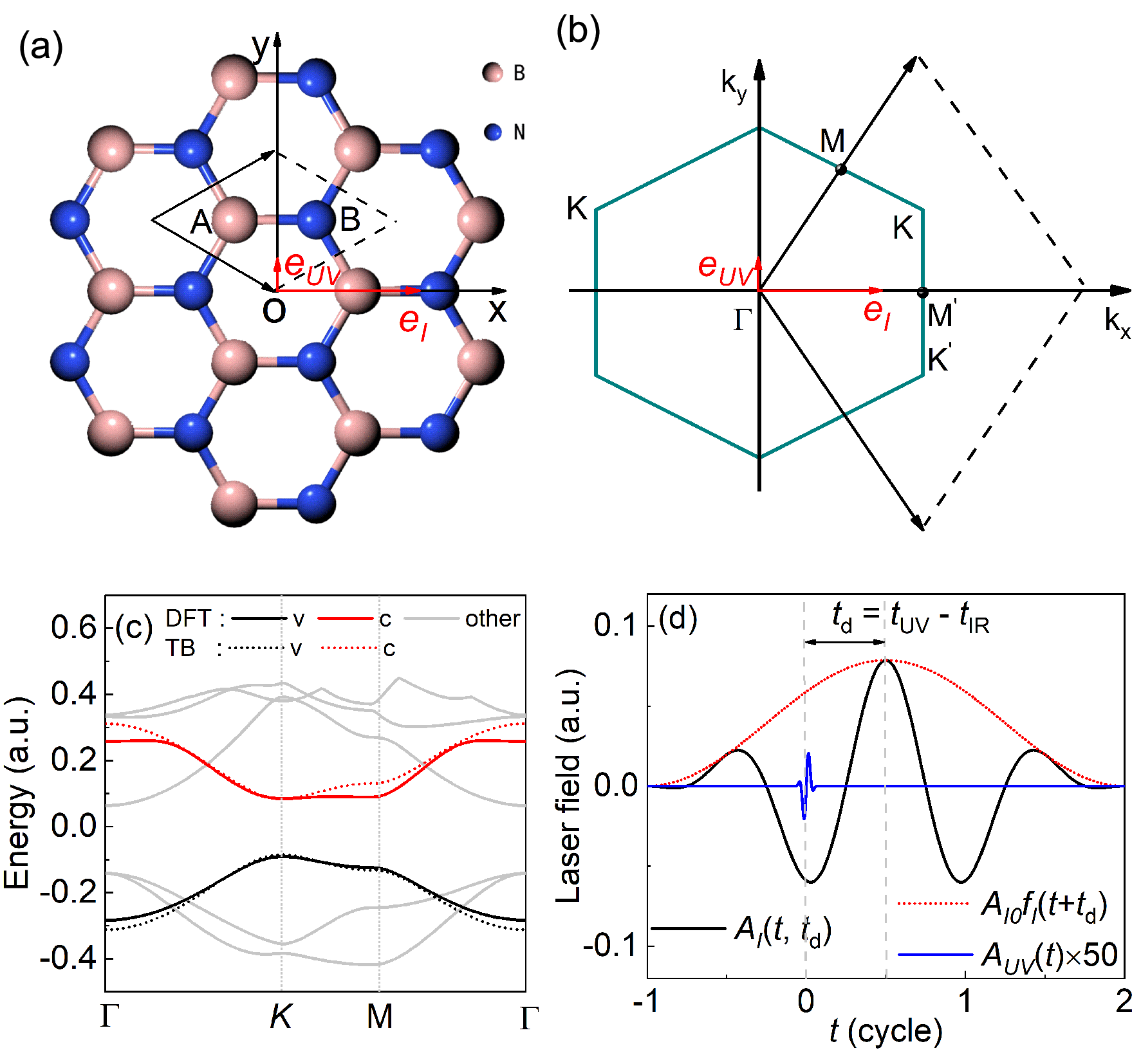}}
\caption{(a) Hexagonal lattice structure of monolayer hBN.
Boron and nitrogen atoms are labeled by ``A" and ``B", respectively.
(b) First Brillouin zone of hBN with high symmetry points $\Gamma$, $\textsc{M}$, $\textsc{M}^{\prime}$, $\textsc{K}$ and $\textsc{K}^{\prime}$.
The unit vectors $\textit{\textbf{e}}_{I}$ and $\textit{\textbf{e}}_{\textrm{UV}}$ corresponding to the polarization of the IR and UV lasers, are oriented along the $x$ and $y$ directions, respectively.
(c) Dispersion relations of energy bands calculated by the DFT and the tight-binding approximation.
(d) Schematic of the time delay between the IR pump laser and the UV probe pulse.
}
\label{fig:graph1}
\end{center}
\end{figure}


Next, we numerically simulate the ATAS of monolayer hBN using TBDMEs in the Houston representation.
Within the dipole approximation, these equations take the form:

\vspace{-0.4cm}
\begin{align}
&i\dfrac{d}{d t}\rho_{mn}(\textrm{\textbf{k}}_t,t,t_{d}) = [ \varepsilon_{mn}(\textrm{\textbf{k}}_t)- i \Gamma_{mn}]\rho_{mn}(\textrm{\textbf{k}}_t,t,t_{d}) + \nonumber\\
&\textit{\textbf{E}}(t, t_{d}) \cdot \sum_{l} [ \textrm{\textbf{D}}^{\textrm{\textbf{k}}_t}_{ml} \rho_{ln}(\textrm{\textbf{k}}_t,t,t_{d}) - \rho_{ml}(\textrm{\textbf{k}}_t,t,t_{d}) \textrm{\textbf{D}}^{\textrm{\textbf{k}}_t}_{ln}],
\end{align}
where $\varepsilon_{mn}(\textbf{k})=\varepsilon_{m}(\textbf{k})-\varepsilon_{n}(\textbf{k})$,
and $m$, $n$ denote $v$ or $c$ band.
Here, the relaxation parameters are set as
$\Gamma_{cv}=\Gamma_{vc} = 0.004$ a.u. $\equiv \Gamma_{0}$ \cite{GioCistaro}, and $\Gamma_{cc}=\Gamma_{vv}=0$.
The transition dipole moment is defined as $\textrm{\textbf{D}}_{mn}^{\textbf{k}} = i \langle u_{m,\textbf{k}}(\textbf{r}) \vert \triangledown_{\textbf{k}} \vert u_{n,\textbf{k}}(\textbf{r}) \rangle$,
where $u_{m,\textbf{k}}(\textbf{r})$ represents the periodic part of the Bloch wavefunction for band $m$.

The crystal quasimomentum is
$\textbf{\textrm{k}}_{t}=\textbf{\textrm{k}} + \textit{\textbf{A}}(t, t_d)$,
where
$\textit{\textbf{A}}(t,t_d)=\textit{\textbf{A}}_{I}(t, t_{d}) +\textit{\textbf{A}}_{\textrm{UV}}(t) = \textit{A}_{I0} f_{I}(t+t_d) \cos(\omega_{I} t+\omega_{I} t_d) \textit{\textbf{e}}_{I} + \textit{A}_{\textrm{UV}} f_{\textrm{UV}}(t) \sin(\omega_{\textrm{UV}}t) \textit{\textbf{e}}_{\textrm{UV}}$
is the total vector potential of the infrared (IR) pump laser and the ultraviolet (UV) probe pulse, as shown in Fig. 1(d).
The envelope functions are given by
$f_{I}(t) = \cos^2 (\omega_{I}t / 2n)$
and
$f_{\textrm{UV}}(t) = e^{-(4 ln 2)(t/ \tau_{\textrm{UV}})^{2}}$ with $n = 3$ and $\tau_{\textrm{UV}} \approx 90$ attoseconds.
The amplitudes $\textit{A}_{I0}$ and $\textit{A}_{\textrm{UV}}$ correspond to laser intensities of $5 \times 10^{10}$ W/cm$^{2}$ and $5 \times 10^{8}$ W/cm$^{2}$, respectively.
The frequencies $\omega_{I}$ and $\omega_{\textrm{UV}}$ correspond to the wavelengths of $3000$ nm and $200$ nm, respectively,
and $T = 2\pi/\omega_{I}$ is the period of the IR laser field.
The unit vectors $\textit{\textbf{e}}_{I}$ and $\textit{\textbf{e}}_{\textrm{UV}}$ point along the $x$ and $y$ directions, as illustrated in Figs. 1(a) and 1(b).
The corresponding electric field is given by $\textit{\textbf{E}}(t) = - \partial \textit{\textbf{A}}(t) / \partial t$.
The time delay is defined as $t_d = t_{\textrm{UV}} - t_{IR}$, where $t_{\textrm{UV}} = 0$, and $t_{IR}$ is the peak of the IR laser envelope, respectively.
When $t_d = 0$, the maxima of the two pulses coincide.

For a given time delay $t_{d}$, the response function of the UV pulse is calculated by \cite{MetteBGaarde}

\vspace{-0.4cm}
\begin{align}
S(\omega,t_{d}) = 2\sum_{\textrm{\textbf{k}}}{\operatorname{Im}[\tilde{\textit{j}}_{\textrm{\textbf{k}}}(\omega,t_{d}) \tilde{\textit{A}}_{\textrm{UV}}^{*}(\omega)]},
\end{align}
where the Fourier-transformed current is defined by $\tilde{\textit{j}}_{\textrm{\textbf{k}}}(\omega,t_{d}) =\int^{\infty}_{-\infty}  \sum_{m,n} \textit{\textbf{e}}_{\textrm{UV}} \cdot \textbf{\textrm{P}}_{mn}^{\textbf{\textrm{k}}_{t}}  \rho_{n m}(\mathbf{k}_{t},t,t_d)  e^{-i \omega t} dt$,
and the momentum matrix elements are given by  $\mathbf{P}_{cc}^{\mathbf{k}} = \triangledown_{\mathbf{k}} \varepsilon_{c}(\mathbf{k}) = -\mathbf{P}_{vv}^{\mathbf{k}}$,
and
$\mathbf{P}_{cv}^{\mathbf{k}} = i (\varepsilon_{c}(\mathbf{k}) - \varepsilon_{v}(\mathbf{k})) \mathbf{D}_{cv}^{\mathbf{k}}$.
Here, $\tilde{\textit{A}}_{\textrm{UV}}^{*}(\omega)$ denotes the complex conjugate of the Fourier transform of $A_{\textrm{UV}}(t)$.

The ATAS can then be calculated as

\vspace{-0.4cm}
\begin{align}
\Delta S(\omega,t_{d}) = S(\omega,t_{d}) - S^{\textrm{UV}}(\omega),
\end{align}
where $S^{\textrm{UV}}(\omega)$ denotes the response function in the absence of the IR laser field.

We also solve the TBDMEs by artificially setting $\boldsymbol{\mathcal{A}} \left(\textbf{\textrm{k}}\right) = \textbf{\textrm{D}}_{cc}^{\textbf{\textrm{k}}} - \textbf{\textrm{D}}_{vv}^{\textbf{\textrm{k}}} = 0$.
In this case, the response function is

\vspace{-0.4cm}
\begin{align}
S^{\boldsymbol{\mathcal{A}} = 0}(\omega,t_{d}) = 2\sum_{\textrm{\textbf{k}}}{\operatorname{Im}[\tilde{\textit{j}}^{\boldsymbol{\mathcal{A}} = 0}_{\textrm{\textbf{k}}}(\omega,t_{d}) \tilde{\textit{A}}_{\textrm{UV}}^{*}(\omega)]},
\end{align}
where,
$\tilde{\textit{j}}^{\boldsymbol{\mathcal{A}} = 0}_{\textrm{\textbf{k}}}(\omega,t_{d})$
is the Fourier transform of the current under the condition $\boldsymbol{\mathcal{A}} = 0$.

The corresponding ATAS is evaluated by

\vspace{-0.4cm}
\begin{subequations}
\begin{align}
\Delta S^{\boldsymbol{\mathcal{A}} = 0}(\omega,t_{d}) =& S^{\boldsymbol{\mathcal{A}} = 0}(\omega,t_{d}) - S^{\textrm{UV}}(\omega),\\
\Delta S^{\boldsymbol{\mathcal{A}}}(\omega,t_{d})=& S(\omega,t_{d})- S^{\boldsymbol{\mathcal{A}}=0}(\omega,t_{d}).
\end{align}
\end{subequations}
Here, the ATAS described by Eq. (5a) is obtained by isolating the influence of the Berry connection and is, in fact, primarily governed by the TDMs.
In contrast, the ATAS in Eq. (5b) is dominated by the Berry connection.

\subsection{Time-dependent density functional theory}

Because our density-matrix equations consider only two energy bands of hBN, we check our main ATAS results using the TDDFT \cite{Ullrich}.
Within the TDDFT framework, the wavefunction evolution is computed by propagating the Kohn-Sham equations \cite{MALMarques,ZNourbakhsh,GuillaumeLB}.
In the underlying density functional theory (DFT) calculation, we employ norm-conserving pseudopotentials and use the generalized gradient approximation with Perdew-Burke-Ernzerhof parameterization to treat the exchange-correlation potential.
As shown in Fig. 1(c), the DFT band structure is plotted with solid lines: the $v$ and $c$ bands are indicated by black and red solid lines, respectively, while the other bands are shown with gray lines.

For the calculation of the harmonic generation in monolayer hBN, a $60 \times 60 \times 1 $ k-point mesh is used to sample the first Brillouin zone.
The vector potential $\textit{\textbf{A}}(t)$ is the same as that used in TBDMEs.
The total electronic current $ \textit{\textbf{j}}(\mathbf{r},t,t_d)$ is computed from time-evolved wavefunctions.
The current can be evaluated using $\tilde{\textit{j}}(\omega,t_d)= \int^{\infty}_{-\infty} [ \textit{\textbf{e}}_{\textrm{UV}} \cdot \int_{\Omega} d^{3} \mathbf{r} \textit{\textbf{j}}(\mathbf{r},t,t_d) ] e^{-i \omega t} dt$, where $\Omega$ is the volume of the physical system.
The OCTOPUS package \cite{Strubbe,Strubbe2} is employed to perform these simulations.

\subsection{Numerical simulation results}

\begin{figure}[t]
\begin{center}
{\includegraphics[width=8.5cm,height=12.0cm]{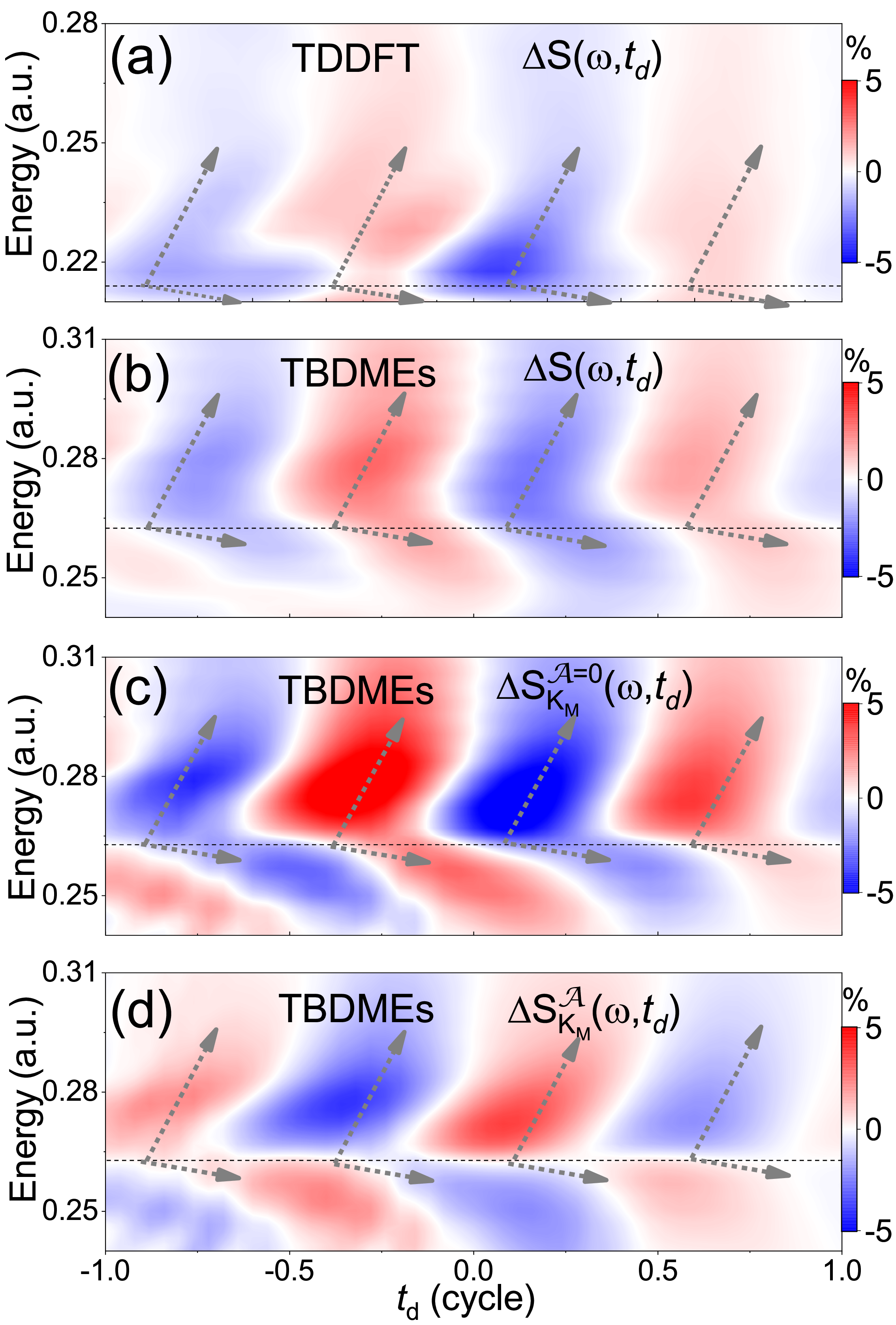}}
\caption{(a) ATAS as a function of the time delay in units of IR laser optical cycles, simulated by the TDDFT.
The black dashed line indicates an energy gap of approximately $0.214$ a.u. at $\textrm{M}$ point, as predicted by DFT.
(b) ATAS simulation calculated by Eq. (3), based on TBDMEs within the tight-binding approximation, where $\varepsilon_{cv}(\textbf{\textrm{k}}_{\textrm{M}}) \approx 0.262$ a.u.
(c), (d) Same as (b), but calculated by Eqs. (5a) and (5b), respectively.
In panels (a)-(d), the gray arrows highlight the fishbone structure in the ATAS.
}
\label{fig:graph1}
\end{center}
\end{figure}


As shown in Fig. 1(c), the energy gap at the $\textrm{M}$ point, calculated using DFT, is approximately $0.214$ a.u.
In Fig. 2(a), this gap energy is indicated by the black horizontal dashed line, around which we present the ATAS of monolayer hBN calculated by Eq. (3), based on the TDDFT simulations.
It should be noted that the spectra have been normalized by $S^{\textrm{UV}}(\omega)$, and the same applies to the following figures.
The fishbone structure highlighted with gray dashed arrows is clearly visible.
Interestingly, this structure exhibits a period of $T$, in contrast to the similar $T/2$-periodic structure in Refs. \cite{Dong4}.

Based on the TBDMEs, the ATAS results calculated by Eq. (3) are shown in Fig. 2(b).
Here, the black dashed line indicates $\varepsilon_{cv}(\textbf{\textrm{k}}_{\textrm{M}}) \approx 0.262$ a.u., obtained from gapped graphene with $\Delta_g = 0.17$ a.u.
It can be seen that the simulation results from the two methods are qualitatively consistent.
Similar to the TDDFT simulation, the fishbone structure marked by gray arrows exhibits a period of $T$.

To isolate the influence of the Berry connection on the ATAS, we calculate the spectra using Eq. (5a), and the corresponding results are shown in Fig. 2(c).
The spectra exhibit a periodic structure similar to that in Fig. 2(b), but with significantly enhanced spectral intensity.
Figure 2(d) presents the spectra calculated using Eq. (5b), which reflect the contribution of the Berry connection $\boldsymbol{\mathcal{A}} \left(\textbf{\textrm{k}}\right)$ on the ATAS.
Notably, at the same time delay $t_d$, the spectra in Fig. 2(c) and Fig. 2(d) display opposite signs.
This phase difference helps explain the enhanced spectral intensity observed in Fig. 2(c) relative to Fig. 2(b).

\section{Analytical study of the ATAS}
\label{s3}

\subsection{Numerical ATAS for single-electron model}

In order to obtain analytical ATAS expressions and reveal the underlying mechanism of the fishbone structure, we analyze the electrons located at the $\textrm{M}$ points.
There are two inequivalent $\textrm{M}$ points in the first Brillouin zone, denoted as $\textrm{M}$ and $\textrm{M}^{\prime}$ as shown in Fig. 1(b).
For $\textrm{M}^{\prime}$, the $y$ component of $\textbf{\textrm{D}}_{cv}^{\textbf{\textrm{k}}_{\textrm{M}^{\prime}}}$ is zero, preventing the transition of the electron from $v$ to $c$ band excited by the UV pulse (see Sec. II of Supplementary Materials for details).
Therefore, we can investigate the ATAS by considering one electron at the $\textrm{M}$ point.

In Fig. 3, we present the ATAS obtained using TBDMEs based on the single-electron model.
The fishbone structure, marked by the gray arrows, is qualitatively consistent with the results shown in Fig. 2(b).
The qualitative agreement between Fig. 2(b) and Fig. 3 justifies the use of the single-electron model to further analyze the ATAS.
In the following, we derive analytical results based on this model to reveal the underlying mechanism.

\subsection{Analytical expressions of the ATAS based on the single-electron model}

Because the UV pulse is relatively short and weak, it can be approximated as a delta function: $\textbf{\textit{E}}_{\textrm{UV}}(t) =\textbf{\textit{A}}_{\textrm{UV}} \delta(t)$.
For $t<0$, the excitation of the electron caused by the IR laser is ignored.
As a result, the time-dependent current is zero, i.e., $j_{\textbf{\textrm{k}}_{\textrm{M}}} (t<0, t_{d}) = 0$.

At $t=0$, the electron is instantaneously excited from the $v$ to $c$ band by the UV pulse.
According to perturbation theory and Eq. (1), the density matrix elements change from the initial values
$\rho_{vv} (\textbf{\textrm{k}}_t,t < 0^{-},t_{d}) = 1$,
$\rho_{cc} (\textbf{\textrm{k}}_t,t < 0^{-},t_{d}) = 0$,
and $\rho_{cv} (\textbf{\textrm{k}}_t,t < 0^{-},t_{d}) = 0$
to $\rho_{vv} (\textbf{\textrm{k}}_t,t = 0^{+},t_{d}) \approx 1$, $\rho_{cc} (\textbf{\textrm{k}}_t,t = 0^{+},t_{d}) \approx 0$, and $\rho_{cv} (\textbf{\textrm{k}}_t,t = 0^{+},t_{d}) \approx -i \textbf{\textit{A}}_{\textrm{UV}} \cdot \textbf{\textrm{D}}_{cv}^{\textbf{\textrm{k}}_{\textrm{M}}}$.

For $t>0$, the time evolution of density-matrix elements is dominated by the IR laser, and the interband excitation of the electron caused by the IR laser is ignored (i.e., assuming $\textbf{\textit{E}}_{I}(t, t_d) \cdot \textrm{\textbf{D}}_{cv}^{\textrm{\textbf{k}}_t} \approx 0$).
Under these assumptions, $\rho_{cv} (\textbf{\textrm{k}}_t,t > 0, t_{d}) = -i \textbf{\textit{A}}_{\textrm{UV}} \cdot \textbf{\textrm{D}}_{cv}^{\textbf{\textrm{k}}_{\textrm{M}}} e^{-i \int^{t}_{0} (\varepsilon_{cv}(\textbf{\textrm{k}}_{t^{\prime}} ) + \textit{\textbf{E}}_{I}(t^{\prime}, t_{d}) \cdot \boldsymbol{\mathcal{A}}(\textbf{\textrm{k}}_{t^{\prime}}) )d t^{\prime}} e^{- \Gamma_{0} t}$.
The resulting time-dependent current is given by

\vspace{-0.4cm}
\begin{align}
j_{\textbf{\textrm{k}}_{\textrm{M}}} (t, t_{d}) = & ( \textbf{\textit{e}}_{\textrm{UV}} \cdot \textbf{\textrm{P}}_{vc}^{\textbf{\textrm{k}}_t} ) \rho_{cv} (\textbf{\textrm{k}}_t, t, t_{d}) +c.c. \nonumber\\
= & - \varepsilon_{cv}(\textbf{\textrm{k}}_{\textrm{M}} ) (\textbf{\textit{e}}_{\textrm{UV}} \cdot \textbf{\textrm{D}}_{vc}^{\textbf{\textrm{k}}_t}) (\textbf{\textit{A}}_{\textrm{UV}} \cdot \textbf{\textrm{D}}_{cv}^{\textbf{\textrm{k}}_{\textrm{M}}}) \nonumber\\
& \exp  \{ -i \int^{t}_{0} \mathcal{C}(t^{\prime}, t_d)  d t^{\prime} \} e^{- \Gamma_{0} t} + c.c.,
\end{align}
where
$\mathcal{C}(t^{\prime}, t_d) = \varepsilon_{cv}\left( \textbf{\textrm{k}}_{\textrm{M}} \right) + E_{I}\left(t^{\prime}, t_d\right) \mathcal{A}_{x}\left(\textbf{\textrm{k}}_{\textrm{M}}\right)$
and $\mathcal{A}_{x} \left(\textbf{\textrm{k}}\right)$
is the $x$ component of the Berry connection $\boldsymbol{\mathcal{A}} \left(\textbf{\textrm{k}}\right)$.
In the derivation of Eq. (6), we have used the conditions of
$\varepsilon_{cv}(\textbf{\textrm{k}}_{t}) = \varepsilon_{cv}(\textbf{\textrm{k}}_{\textrm{M}} )$ and
$\mathcal{A}_{x}\left(\textbf{\textrm{k}}_{t}\right) = \mathcal{A}_{x}\left(\textbf{\textrm{k}}_{\textrm{M}}\right) $ (see Sec. II of the Supplemental Material for details).

\begin{figure}[t]
\begin{center}
{\includegraphics[width=8.5cm,height=4.5cm]{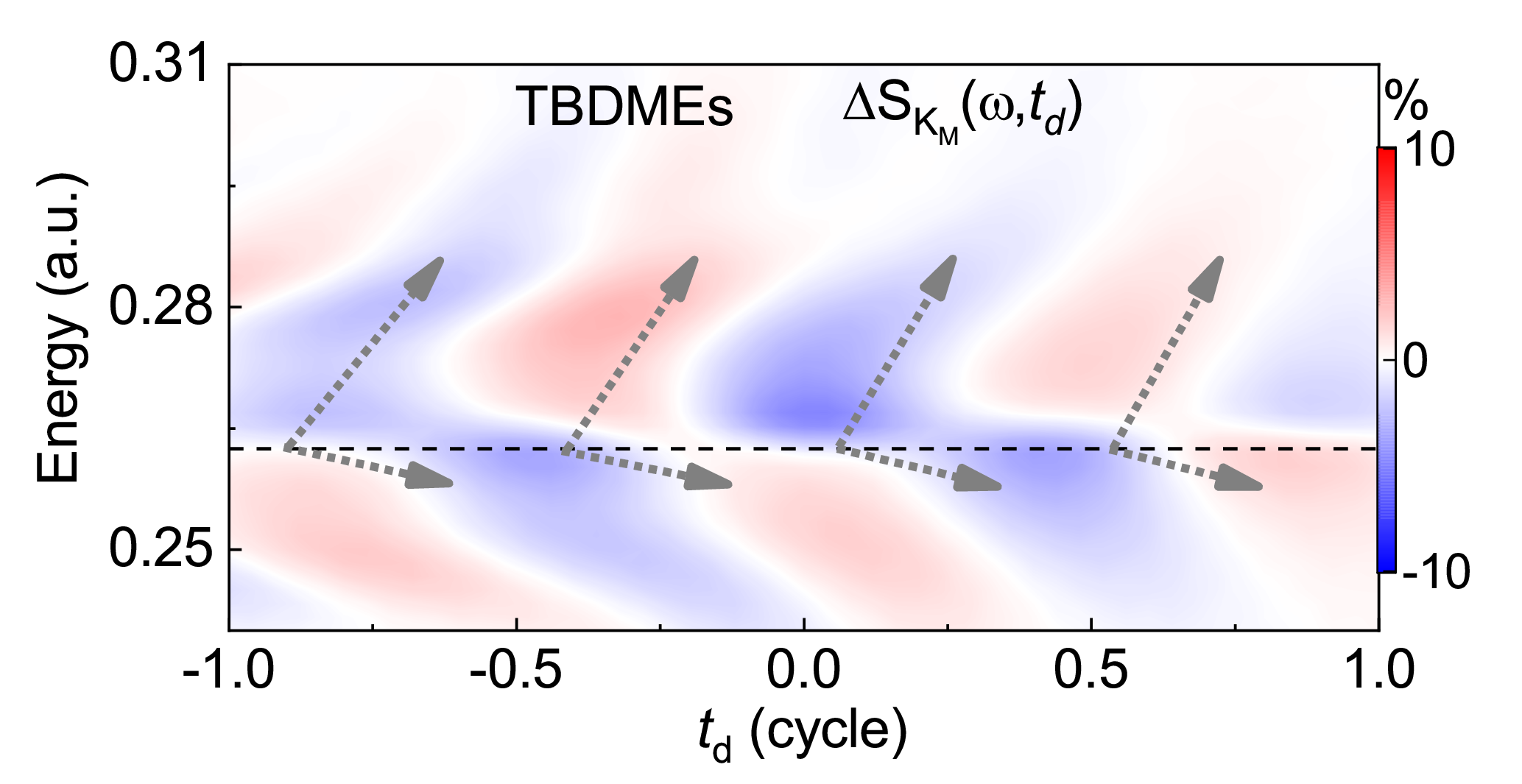}}
\caption{Numerical ATAS calculated using the single-electron model, considering only the electron located at $\textrm{M}$ point.
}
\label{fig:graph1}
\end{center}
\end{figure}

With the IR laser turned off, the current is
$\textit{j}_{\textbf{\textrm{k}}_{\textrm{M}}}^{\textrm{UV}}(t) = -2 \varepsilon_{c v}({\textbf{\textrm{k}}_{\textrm{M}}}) A_{\textrm{UV}} | \textrm{D}_{c v, y}^{{\textbf{\textrm{k}}_{\textrm{M}}}} |^2 \cos (\int_0^t \varepsilon_{c v}({\textbf{\textrm{k}}_{\textrm{M}}}) d t^{\prime}) e^{-\Gamma_0 t}$ for $t > 0$.
The corresponding response function is

\vspace{-0.4cm}
\begin{align}
S^{\textrm{UV}}_{\textbf{\textrm{k}}_{\textrm{M}}}  (\omega) & \propto - \operatorname{Re}\left[ \tilde{\textit{j}}_{\textbf{\textrm{k}}_{\textrm{M}}}^{\textrm{UV}}\left(\omega\right)\right] = - \operatorname{Re}\left[\int_{0}^{\infty} \textit{j}_{\textbf{\textrm{k}}_{\textrm{M}}}^{\textrm{UV}}(t) e^{-i \omega t} d t\right] \nonumber\\
&= \boldsymbol{\mathcal{F}}_0 L(\omega,\varepsilon_{c v}({\textbf{\textrm{k}}_{\textrm{M}}})),
\end{align}
where
$\boldsymbol{\mathcal{F}}_0 = \varepsilon_{c v}({\textbf{\textrm{k}}_{\textrm{M}}}) A_{\textrm{UV}}|\textrm{D}_{c v, y}^{{\textbf{\textrm{k}}_{\textrm{M}}}} |^2$,
and $L(\omega,x) = \frac{\Gamma_{0}}{\Gamma_{0}^{2}+(\omega-x)^{2}}$ is the Lorentzian line shape centered at $x$.

When the IR laser is turned on, the term $\textbf{\textit{e}}_{\textrm{UV}} \cdot \textbf{\textrm{D}}_{vc}^{\textbf{\textrm{k}}_t}$ in Eq. (6) can be approximated by a Taylor expansion around $\textbf{\textrm{k}}_{\textrm{M}}$, i.e.,
$\textrm{D}_{vc,y}^{\textbf{\textrm{k}}_t} = \operatorname{Re} [\textrm{D}_{vc,y}^{\textbf{\textrm{k}}_t}] + i \operatorname{Im} [\textrm{D}_{vc,y}^{\textbf{\textrm{k}}_t}]
\approx
\operatorname{Re} [\textrm{D}_{vc,y}^{\textbf{\textrm{k}}_{\textrm{M}}}]
+
\frac{1}{2} \nabla_{\textrm{k}_x}^2 \operatorname{Re} [\textrm{D}_{vc,y}^{\textbf{\textrm{k}}_{\textrm{M}}}] \textit{A}_{I0}^{2} f_{I}^{2}(t_d) \cos^{2}(\omega_{I} t+\omega_{I} t_d) + i \nabla_{\textrm{k}_x} \operatorname{Im} [\textrm{D}_{vc,y}^{\textbf{\textrm{k}}_{\textrm{M}}}] \textit{A}_{I0} f_{I}(t_d) \cos(\omega_{I} t+\omega_{I} t_d)$.
Here, we retain only the leading second-order terms in the expansion $\operatorname{Re} [\textrm{D}_{vc,y}^{\textbf{\textrm{k}}_t}] $ and $\operatorname{Im} [\textrm{D}_{vc,y}^{\textbf{\textrm{k}}_t}]$ (see Sec. I of the Supplemental Material for details).
The current in Eq. (6) can thus be approximated as

\vspace{-0.4cm}
\begin{align}
\textit{j}_{\textbf{\textrm{k}}_{\textrm{M}}} & (t, t_{d})
\approx 2 \boldsymbol{\mathcal{F}}_1 \cos (\int_0^t \mathcal{C}(t^{\prime}, t_d) d t^{\prime}) e^{- \Gamma_{0} t} \nonumber\\
& + 2 \boldsymbol{\mathcal{F}}_2 \sin (\int_0^t \mathcal{C}(t^{\prime}, t_d) d t^{\prime}+ \omega_{I} t + \omega_{I} t_d ) e^{- \Gamma_{0} t}  \nonumber\\
& + 2 \boldsymbol{\mathcal{F}}_2 \sin (\int_0^t \mathcal{C}(t^{\prime}, t_d) d t^{\prime}- \omega_{I} t - \omega_{I} t_d ) e^{- \Gamma_{0} t},
\end{align}
where,
$\boldsymbol{\mathcal{F}}_1 =  - \frac{1}{4}  A_{\textrm{UV}}  A_{I0}^{2} f_{I}^{2}(t_d) \varepsilon_{cv}(\textbf{\textrm{k}}_{\textrm{M}}) \operatorname{Re} [\textrm{D}_{vc,y}^{\textbf{\textrm{k}}_{\textrm{M}}}] \cdot \nabla_{\textrm{k}_x}^2 \operatorname{Re} [\textrm{D}_{vc,y}^{\textbf{\textrm{k}}_{\textrm{M}}}] $,
and
$\boldsymbol{\mathcal{F}}_2 = \frac{1}{2} A_{\textrm{UV}}  A_{I0} f_{I}(t_d) \varepsilon_{cv}(\textbf{\textrm{k}}_{\textrm{M}}) \cdot \operatorname{Re} [\textrm{D}_{vc,y}^{\textbf{\textrm{k}}_{\textrm{M}}}] \nabla_{\textrm{k}_x} \operatorname{Im} [\textrm{D}_{vc,y}^{\textbf{\textrm{k}}_{\textrm{M}}}]$
are coefficients determined by the interband TDMs.
At $t_d = 0$, the amplitudes of $\boldsymbol{\mathcal{F}}_1$ and $\boldsymbol{\mathcal{F}}_2$ are approximately $0.025 \boldsymbol{\mathcal{F}}_0$ and $0.103 \boldsymbol{\mathcal{F}}_0$, respectively.
A straightforward analysis reveals that the increase from a single term in Eq. (6) to three distinct terms in Eq. (8) arises directly from the Taylor expansion of $\textrm{D}_{vc,y}^{\textbf{\textrm{k}}_t}$, which account for the variation of the interband TDMs as the electron undergoes Bloch oscillations in reciprocal space.

The ATAS at $\textrm{M}$ point is given by

\vspace{-0.4cm}
\begin{align}
\Delta & S_{\textbf{\textrm{k}}_{\textrm{M}}} (\omega,  t_{d}) = S_{\textbf{\textrm{k}}_{\textrm{M}}} (\omega, t_{d}) - S^{\textrm{UV}}_{\textbf{\textrm{k}}_{\textrm{M}}}  (\omega) \nonumber\\
& \propto - \operatorname{Re}[ \tilde{\textit{j}}_{\textbf{\textrm{k}}_{\textrm{M}}}(\omega,t_{d}) ] - S^{\textrm{UV}}_{\textbf{\textrm{k}}_{\textrm{M}}}  (\omega) \nonumber\\
& = \Delta S_{\textbf{\textrm{k}}_{\textrm{M}}}^{(0)} (\omega, t_{d})
+ \Delta S_{\textbf{\textrm{k}}_{\textrm{M}}}^{(1)} (\omega, t_{d})
+ \Delta S_{\textbf{\textrm{k}}_{\textrm{M}}}^{(2)} (\omega, t_{d}).
\end{align}
Here, $\Delta S_{\textbf{\textrm{k}}_{\textrm{M}}}^{(0)} (\omega, t_{d})$, $\Delta S_{\textbf{\textrm{k}}_{\textrm{M}}}^{(1)} (\omega, t_{d})$, and $\Delta S_{\textbf{\textrm{k}}_{\textrm{M}}}^{(2)} (\omega, t_{d})$ represent the zeroth-, first-, and second-order structure of the ATAS, evaluated as

\vspace{-0.4cm}
\begin{align}
\Delta S_{\textbf{\textrm{k}}_{\textrm{M}}}^{(0)} (\omega, t_{d})
= - \boldsymbol{\mathcal{F}}_1 J_{0}(c) L(\omega, \varepsilon_{cv}\left( \textbf{\textrm{k}}_{\textrm{M}} \right)),
\end{align}

\vspace{-0.4cm}
\begin{align}
\Delta & S_{\textbf{\textrm{k}}_{\textrm{M}}}^{(1)} (\omega, t_{d}) \approx \nonumber\\
& -  \left[ \boldsymbol{\mathcal{F}}_2 J_{0}(c) - \boldsymbol{\mathcal{F}}_0 J_{1}(c)   \right]  L(\omega, \varepsilon_{cv}\left( \textbf{\textrm{k}}_{\textrm{M}} \right) + \omega_{I})\sin (\omega_{I} t_d)\nonumber\\
&+ \left[ \boldsymbol{\mathcal{F}}_2 J_{0}(c) - \boldsymbol{\mathcal{F}}_0 J_{1}(c)   \right] F(\omega, \varepsilon_{cv}\left( \textbf{\textrm{k}}_{\textrm{M}} \right) + \omega_{I})\cos (\omega_{I} t_d) \nonumber\\
& + \left[ \boldsymbol{\mathcal{F}}_2 J_{0}(c) - \boldsymbol{\mathcal{F}}_0 J_{1}(c)   \right] L(\omega, \varepsilon_{cv}\left( \textbf{\textrm{k}}_{\textrm{M}} \right) -\omega_{I})\sin (\omega_{I} t_d)\nonumber\\
&+ \left[ \boldsymbol{\mathcal{F}}_2 J_{0}(c) - \boldsymbol{\mathcal{F}}_0 J_{1}(c)   \right] F(\omega, \varepsilon_{cv}\left( \textbf{\textrm{k}}_{\textrm{M}} \right) -\omega_{I})\cos (\omega_{I} t_d),
\end{align}
and

\vspace{-0.4cm}
\begin{align}
\Delta S_{\textbf{\textrm{k}}_{\textrm{M}}}^{(2)} & (\omega, t_{d}) =   \boldsymbol{\mathcal{F}}_2 J_{1}(c) L(\omega, \varepsilon_{cv}\left( \textbf{\textrm{k}}_{\textrm{M}} \right) + 2\omega_{I})\cos (2\omega_{I} t_d)  \nonumber\\
& +  \boldsymbol{\mathcal{F}}_2 J_{1}(c) F(\omega, \varepsilon_{cv}\left( \textbf{\textrm{k}}_{\textrm{M}} \right) + 2\omega_{I})\sin (2\omega_{I} t_d) \nonumber\\
& +  \boldsymbol{\mathcal{F}}_2 J_{1}(c) L(\omega, \varepsilon_{cv}\left( \textbf{\textrm{k}}_{\textrm{M}} \right) -2\omega_I)\cos (2\omega_{I} t_d)  \nonumber\\
& -  \boldsymbol{\mathcal{F}}_2 J_{1}(c) F(\omega, \varepsilon_{cv}\left( \textbf{\textrm{k}}_{\textrm{M}} \right) -2\omega_I )\sin (2\omega_{I} t_d).
\end{align}
In the process of deducing Eq. (11), the small terms involving $\boldsymbol{\mathcal{F}}_1 J_{1}(c)$ have been ignored.
$F(\omega,x) = \dfrac{\omega - x}{\Gamma_{0}^{2} + (\omega - x)^{2}}$ denotes the Fano line shape centered at $x$.
Here, $c = \mathcal{A}_{x}\left(\textbf{\textrm{k}}_{\textrm{M}}\right) A_{I0} f_I(t_d)$, and $J_{n}(x)$ denotes the $n$th-order Bessel function. Higher-order terms with $n \geq 2$ are ignored.
At $t_d = 0$, $c=0.1367$, $J_{0}(c) \approx 0.995$, and $J_{1}(c) \approx 0.068$.

\begin{figure}[t]
\begin{center}
{\includegraphics[width=8.5cm,height=10cm]{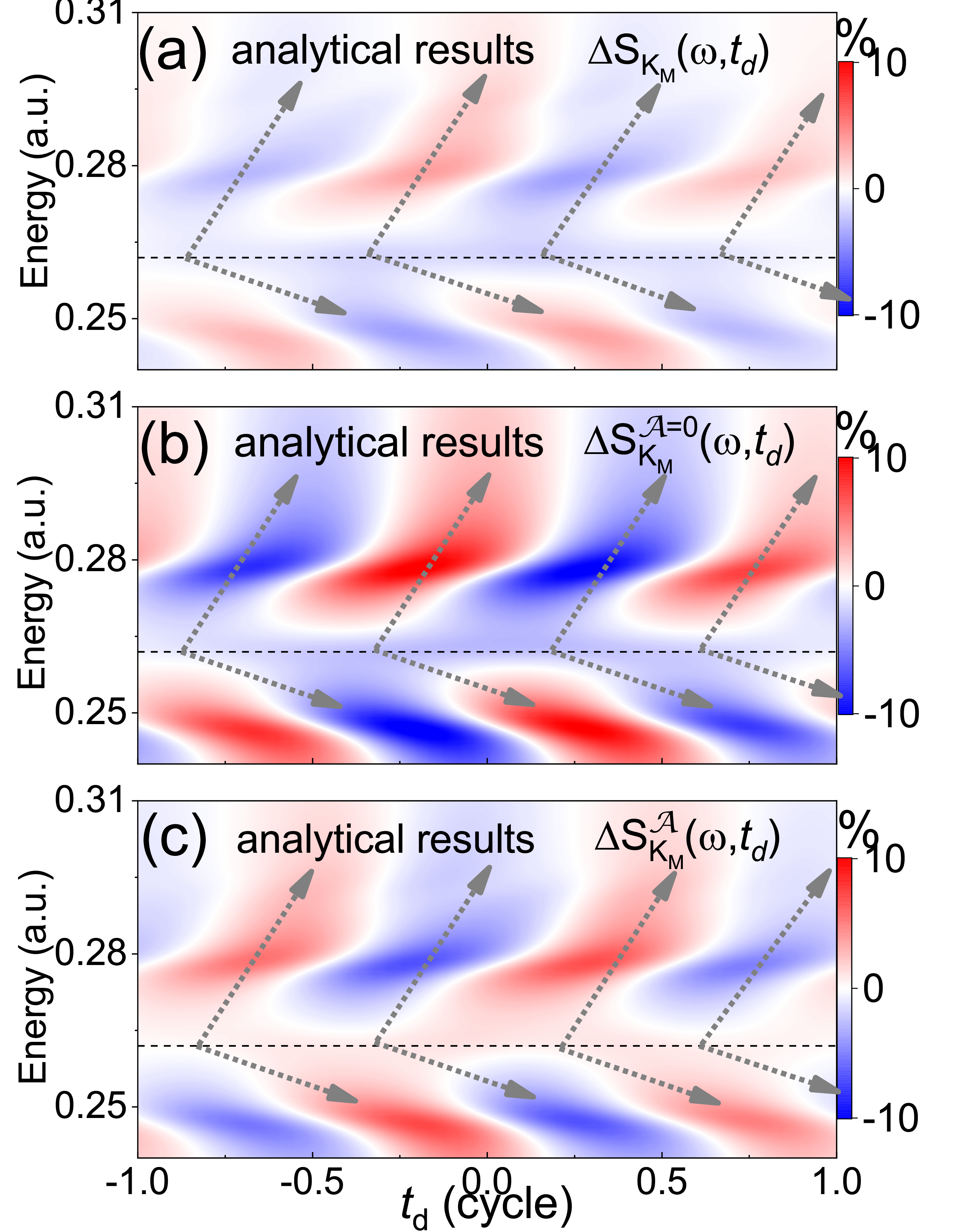}}
\caption{(a) Analytical spectra $\Delta S_{\textbf{\textrm{k}}_{\textrm{M}}} (\omega, t_{d})$ calculated using Eq. (9).
The horizontal dashed lines indicate the energy $\varepsilon_{c v}({\textbf{\textrm{k}}_{\textrm{M}}})$.
(b) Analytical spectra $\Delta S^{\mathcal{A}=0}_{\textbf{\textrm{k}}_{\textrm{M}}} (\omega, t_{d})$ calculated using Eq. (13).
(c) Analytical spectra $\Delta S^{\mathcal{A}}_{\textbf{\textrm{k}}_{\textrm{M}}} (\omega, t_{d})$ calculated using Eq. (14).
These analytical results are obtained based on the single-electron model.
}
\label{fig:graph1}
\end{center}
\end{figure}

From the derivation, it can be found that the appearance of the higher-order terms in Eq. (9) arises from the oscillation term $\mathcal{A}_{x}\left(\textbf{\textrm{k}}_{\textrm{M}}\right) A_{I}\left(t, t_d\right)$ in the integral $\int_0^t \mathcal{C}(t^{\prime}, t_d) d t^{\prime}$.
By analyzing the analytical ATAS expressions in Eqs. (9)-(12), we conclude that both the interband TDMs $\textrm{D}_{vc,y}^{\textbf{\textrm{k}}_t}$ and the Berry connection component $\mathcal{A}_{x}\left(\textbf{\textrm{k}}_{t}\right)$ play dominant roles in determining the fishbone structure in the ATAS of hBN.

To isolate the influence of the Berry connection on the ATAS, we artificially set $\mathcal{A}_{x}\left(\textbf{\textrm{k}}_{\textrm{M}}\right) = 0$, i.e., $c=0$ in Eqs. (10)-(12).
In this case, the ATAS in Eq. (9) simplifies to

\vspace{-0.4cm}
\begin{align}
\Delta S_{\textbf{\textrm{k}}_{\textrm{M}}}^{\mathcal{A}=0} (\omega, t_{d}) &  = \Delta S_{\textbf{\textrm{k}}_{\textrm{M}}}^{(0) \mathcal{A}=0} (\omega, t_{d}) + \Delta S_{\textbf{\textrm{k}}_{\textrm{M}}}^{(1) \mathcal{A}=0} (\omega, t_{d}) \nonumber\\
& \approx - \boldsymbol{\mathcal{F}}_1 L(\omega, \varepsilon_{cv}(\textbf{\textrm{k}}_{\textrm{M}}) ) \nonumber\\
& - \boldsymbol{\mathcal{F}}_2 L(\omega, \varepsilon_{cv}(\textbf{\textrm{k}}_{\textrm{M}}) + \omega_{I})\sin (\omega_{I} t_d) \nonumber\\
&+ \boldsymbol{\mathcal{F}}_2 F(\omega, \varepsilon_{cv}(\textbf{\textrm{k}}_{\textrm{M}}) + \omega_{I})\cos (\omega_{I} t_d) \nonumber\\
&+ \boldsymbol{\mathcal{F}}_2 L(\omega, \varepsilon_{cv}(\textbf{\textrm{k}}_{\textrm{M}}) - \omega_{I})\sin (\omega_{I} t_d) \nonumber\\
&+ \boldsymbol{\mathcal{F}}_2 F(\omega, \varepsilon_{cv}(\textbf{\textrm{k}}_{\textrm{M}}) - \omega_{I})\cos (\omega_{I} t_d).
\end{align}

\begin{figure*}[t]
\begin{center}
{\includegraphics[width=18cm,height=8cm]{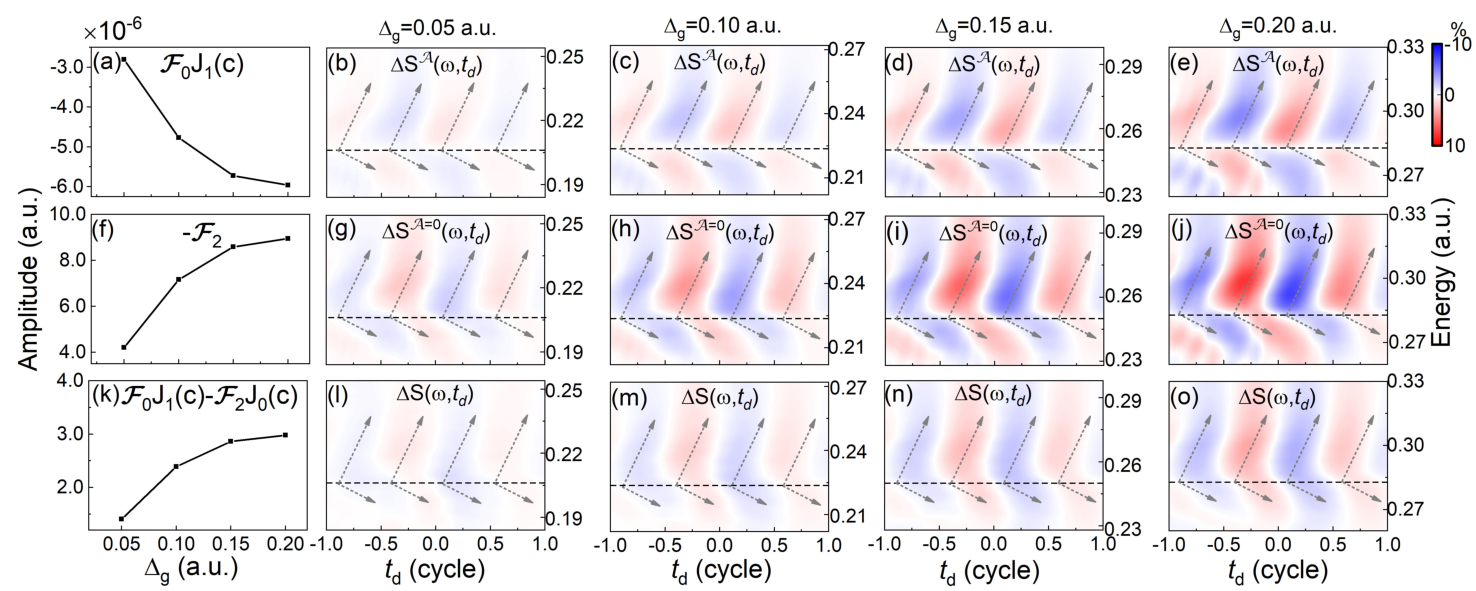}}
\caption{(a) Analytical coefficient $\boldsymbol{\mathcal{F}}_0 J_{1}(c)$ at $t_d =0$ as a function of the gap energy.
(b)-(e) Numerical spectra $\Delta S^{\mathcal{A}} (\omega, t_{d})$ calculated using Eq. (5b) for gap energies of $\Delta_g = 0.05$ a.u., 0.1 a.u., 0.15 a.u., and 0.20 a.u., respectively.
(f) Same as (a), but for the analytical coefficient $- \boldsymbol{\mathcal{F}}_2$.
(g)-(j) Same as (b)-(e), but for the numerical spectra $\Delta S^{\mathcal{A}=0} (\omega, t_{d})$ calculated using Eq. (5a).
(k) Same as (a), but for the analytical coefficient $\boldsymbol{\mathcal{F}}_0 J_{1}(c) - \boldsymbol{\mathcal{F}}_2 J_{0}(c)$.
(l)-(o) Same as (b)-(e), but for the numerical spectra $\Delta S (\omega, t_{d})$ calculated using Eq. (3).
}
\label{fig:graph1}
\end{center}
\end{figure*}

In addition, the contribution of the Berry connection to the ATAS can be evaluated by

\vspace{-0.4cm}
\begin{align}
\Delta S_{\textbf{\textrm{k}}_{\textrm{M}}}^{\mathcal{A}} & (\omega, t_{d}) = \Delta S_{\textbf{\textrm{k}}_{\textrm{M}}} (\omega, t_{d}) - \Delta S_{\textbf{\textrm{k}}_{\textrm{M}}}^{\mathcal{A}=0} (\omega, t_{d})  \nonumber\\
\approx & \boldsymbol{\mathcal{F}}_0 J_{1}(c)    L(\omega, \varepsilon_{cv}\left( \textbf{\textrm{k}}_{\textrm{M}} \right) + \omega_{I})\sin (\omega_{I} t_d)\nonumber\\
& - \boldsymbol{\mathcal{F}}_0 J_{1}(c)    F(\omega, \varepsilon_{cv}\left( \textbf{\textrm{k}}_{\textrm{M}} \right) + \omega_{I})\cos (\omega_{I} t_d) \nonumber\\
& - \boldsymbol{\mathcal{F}}_0 J_{1}(c)    L(\omega, \varepsilon_{cv}\left( \textbf{\textrm{k}}_{\textrm{M}} \right) -\omega_{I})\sin (\omega_{I} t_d)\nonumber\\
& - \boldsymbol{\mathcal{F}}_0 J_{1}(c)   F(\omega, \varepsilon_{cv}\left( \textbf{\textrm{k}}_{\textrm{M}} \right) -\omega_{I})\cos (\omega_{I} t_d).
\end{align}
In the process of deducing Eq. (14), we have considered $1 - J_{0}(c) \approx 0$ and $\boldsymbol{\mathcal{F}}_2 J_{1}(c) \approx 0$.

\subsection{Analytical results of ATAS}

Figure 4(a) presents the analytical spectra $\Delta S_{\textbf{\textrm{k}}_{\textrm{M}}} (\omega, t_{d})$ calculated using Eq. (9).
As shown in Eqs. (10)-(12), the spectra $\Delta S_{\textbf{\textrm{k}}_{\textrm{M}}}^{(0)} (\omega, t_{d})$, $\Delta S_{\textbf{\textrm{k}}_{\textrm{M}}}^{(1)} (\omega, t_{d})$, and $\Delta S_{\textbf{\textrm{k}}_{\textrm{M}}}^{(2)} (\omega, t_{d})$ are centered at $\varepsilon_{cv}\left( \textbf{\textrm{k}}_{\textrm{M}} \right)$ labeled by the black lines, $\varepsilon_{cv}\left( \textbf{\textrm{k}}_{\textrm{M}} \right) \pm \omega_{I}$, and $\varepsilon_{cv}\left( \textbf{\textrm{k}}_{\textrm{M}} \right) \pm 2\omega_{I}$, respectively.
Due to $\vert\boldsymbol{\mathcal{F}}_0 J_{1}(c) - \boldsymbol{\mathcal{F}}_2 J_{0}(c)\vert > \vert\boldsymbol{\mathcal{F}}_1 J_{0}(c)\vert \gg \vert\boldsymbol{\mathcal{F}}_2 J_{1}(c)\vert$, the first-order term $\Delta S_{\textbf{\textrm{k}}_{\textrm{M}}}^{(1)} (\omega, t_{d})$, oscillating with a period of $T$ (or a frequency of $\omega_I$), dominates the ATAS.
The corresponding spectra, marked by the gray arrows, exhibit a fishbone structure that is qualitatively consistent with that observed in Fig. 2(b).
In contrast, the spectra $\Delta S_{\textbf{\textrm{k}}_{\textrm{M}}}^{(0)} (\omega, t_{d})$ and $\Delta S_{\textbf{\textrm{k}}_{\textrm{M}}}^{(2)} (\omega, t_{d})$, oscillating with periods of $0$ and $T/2$ (or frequencies $0$ and $2 \omega_{I}$), respectively, play small roles in the ATAS of Fig. 4(a).

Figure 4(b) presents the analytical spectra $\Delta S_{\textbf{\textrm{k}}_{\textrm{M}}}^{\mathcal{A}=0} (\omega, t_{d})$ calculated using Eq. (13), which are solely contributed by the interband TDMs.
The spectra are still dominated by the first-order term $\Delta S_{\textbf{\textrm{k}}_{\textrm{M}}}^{(1) \mathcal{A}=0} (\omega, t_{d})$, due to $\boldsymbol{\mathcal{F}}_2 \gg \boldsymbol{\mathcal{F}}_1$, as also evidenced in Fig. 2(c).
Moreover, comparing Eq. (13) with Eqs. (10) and (11), since $\boldsymbol{\mathcal{F}}_1 > \boldsymbol{\mathcal{F}}_1 J_{0}(c)$ and $\boldsymbol{\mathcal{F}}_2 \gg \boldsymbol{\mathcal{F}}_2 J_{0}(c) - \boldsymbol{\mathcal{F}}_0 J_{1}(c)$, the spectral amplitude  is significantly enhanced compared to Fig. 4(a).
This enhancement is clearly evident in the comparisons between Figs. 2(b) and 2(c).

Figure 4(c) shows the spectra $\Delta S_{\textbf{\textrm{k}}_{\textrm{M}}}^{\mathcal{A}} (\omega, t_{d})$ calculated using Eq. (14), which arise from the contribution of the Berry connection $\boldsymbol{\mathcal{A}} \left(\textbf{\textrm{k}}\right)$ on the ATAS.

\subsection{Dependence of the ATAS on the gap energy}

Utilizing the tight-binding model, we investigate the dependence of the ATAS on the gap energy.
In Figs. 5(a), 5(f), and 5(k), we show the analytical terms $\boldsymbol{\mathcal{F}}_0 J_{1}(c)$, $- \boldsymbol{\mathcal{F}}_2$, and $\boldsymbol{\mathcal{F}}_0 J_{1}(c) - \boldsymbol{\mathcal{F}}_2 J_{0}(c)$ as a function of the gap energy, which correspond to the coefficients of Eqs. (14), (13), and (11), respectively.
In Figs. 5(b-e), 5(g-j), and 5(l-o), we present numerical spectra $\Delta S^{\mathcal{A}} (\omega, t_{d})$, $\Delta S^{\mathcal{A}=0} (\omega, t_{d})$, and $\Delta S (\omega, t_{d})$, respectively.

It can be observed that as the gap energy increases, the amplitudes of analytical terms $\vert \boldsymbol{\mathcal{F}}_0 J_{1}(c) \vert$, $\vert-\boldsymbol{\mathcal{F}}_2\vert$, and $\vert \boldsymbol{\mathcal{F}}_0 J_{1}(c) - \boldsymbol{\mathcal{F}}_2 J_{0}(c) \vert$  all increase.
Correspondingly, the spectral intensities of $\Delta S^{\mathcal{A}} (\omega, t_{d})$, $\Delta S^{\mathcal{A}=0} (\omega, t_{d})$, and $\Delta S (\omega, t_{d})$ is enhanced in the same tend.
For a fixed gap energy, the amplitudes satisfy the relation $\vert -\boldsymbol{\mathcal{F}}_2 \vert > \vert \boldsymbol{\mathcal{F}}_0 J_{1}(c) \vert > \vert\boldsymbol{\mathcal{F}}_0 J_{1}(c)-\boldsymbol{\mathcal{F}}_2 J_{0}(c)\vert$.
The corresponding intensities of three spectra $\Delta S^{\mathcal{A}=0} (\omega, t_{d})$, $\Delta S^{\mathcal{A}} (\omega, t_{d})$, $\Delta S (\omega, t_{d})$ decrease in the same order.
Moreover, the sign of $\boldsymbol{\mathcal{F}}_0 J_{1}(c)$ is opposite to those of $-\boldsymbol{\mathcal{F}}_2$ and $\boldsymbol{\mathcal{F}}_0 J_{1}(c)-\boldsymbol{\mathcal{F}}_2 J_{0}(c)$, and the coefficients satisfy the relation $\boldsymbol{\mathcal{F}}_0 J_{1}(c) -\boldsymbol{\mathcal{F}}_2 \approx \boldsymbol{\mathcal{F}}_0 J_{1}(c)-\boldsymbol{\mathcal{F}}_2 J_{0}(c)$.
This sign inversion in the analytical terms is directly reflected in the spectral features: the sign of $\Delta S^{\mathcal{A}} (\omega, t_{d})$ is reversed relative to $\Delta S^{\mathcal{A}=0} (\omega, t_{d})$ and $\Delta S (\omega, t_{d})$ at a fixed $(\omega, t_{d})$.
The coefficient relation corresponds to $\Delta S^{\mathcal{A}} (\omega, t_{d}) + \Delta S^{\mathcal{A}=0} (\omega, t_{d}) = \Delta S (\omega, t_{d})$.

\section{Conclusion}
\label{s4}

We calculate the ATAS of hBN using the TDDFT and the TBDMEs within the tight-binding approximation.
The simulation results obtained by both methods exhibit a fishbone structure around the energy gap at the $\textrm{M}$ point, with a temporal period equal to that of the pump laser.
To further investigate this structure, we employ a single-electron model at the $\textsc{M}$ point to derive analytical expressions for the ATAS.
Our analytical results reveal that both the interband TDMs and the Berry connection play important roles in the fishbone structure of the ATAS.
By investigating the dependence of the ATAS on the gap energy, we further reveal the qualitative consistency between our analytical predictions and numerical simulation.
Our study may shed light on the generation mechanism of ATAS in hBN.

\section*{ACKNOWLEDGMENTS}

This work is supported by NSAF (Grant No. 12404394), Hebei Province Optoelectronic Information Materials Laboratory Performance Subsidy Fund Project (No. 22567634H), the High-Performance Computing Center of Hebei University.

\end{document}